\documentclass[12pt]{iopart}

\pdfoutput=1
\usepackage{iopams}
\usepackage{graphicx}
\usepackage{epsfig}
\usepackage[latin1]{inputenc}
\usepackage{amssymb}

\begin{document}

\title{Particle Ratios and the QCD Critical Temperature }
\author{J. Noronha-Hostler$^1$
\footnote{E-mail: hostler@th.physik.uni-frankfurt.de},
Jorge Noronha$^2$
\footnote{E-mail: noronha@phys.columbia.edu},
Carsten Greiner$^1$
\footnote{E-mail: carsten.greiner@th.physik.uni-frankfurt.de}
}

\address{$^1$Institut f\"ur Theoretische Physik, Johann Wolfgang 
Goethe-Universit\"at Frankfurt, Max-von-Laue-Str.1, 
D-60438 Frankfurt am Main, Germany\\
$^2$Department of Physics, Columbia University, 538 West 120$^{th}$ Street, New York,
NY 10027, USA\\[0.2ex]
}

\begin{abstract}
We show how the measured particle ratios at RHIC can be used to provide non-trivial information about the critical temperature of the QCD phase transition. This is obtained by including the effects of highly massive Hagedorn resonances on statistical models, which are used to describe hadronic yields. Hagedorn states are relevant close to $T_c$ and have been shown to decrease $\eta/s$ to the KSS limit and allow for quick chemical equilibrium times in dynamical calculations of hadrons.   The inclusion of Hagedorn states creates a dependence of the thermal fits on the Hagedorn temperature, $T_H$, which is assumed to be equal to $T_c$, and leads to an overall improvement of thermal fits. We find that for Au+Au collisions at RHIC at $\sqrt{s_{NN}}=200$ GeV the best square fit measure, $\chi^2$, occurs at $T_c \sim 176$ MeV and produces a chemical freeze-out temperature of $170.4$ MeV and a baryon chemical potential of $27.8$ MeV.
\end{abstract}
% 
% \pacs{25.75.-q, 12.38.Mh, 05.60.-k, 24.10.Lx}
% %25.75.-q       Relativistic heavy-ion collisions
% %12.38.Mh       Quark-Gluon plasma
% %05.60.-k       Transport processes
% %24.10.Lx       Monte Carlo simulations
% \maketitle

\section{Introduction}\label{intro}

Hagedorn states, heavy resonances that follow an exponential mass spectrum, have been recently shown to play an important role close to the QCD critical temperature, $T_c$.  They have been used to find a low $\eta/s$ in the hadron gas phase \cite{NoronhaHostler:2008ju}, which nears the string theory bound $\eta/s=1/(4\pi)$ \cite{Kovtun:2004de}. Calculations of the trace anomaly including Hagedorn states also fit recent lattice results well and correctly describe the minimum  $c_s^2$ near the phase transition found on the lattice \cite{NoronhaHostler:2008ju}. Furthermore, estimates for the bulk viscosity including Hagedorn states in the hadron gas phase indicate that $\zeta/s$ increases near $T_c$ \cite{NoronhaHostler:2008ju}.

Recent calculations using Hagedorn states to drive hadrons into chemical equilibrium indicate that hadrons do not need to be born in equilibrium at RHIC \cite{Greiner:2004vm,NoronhaHostler:2007jf,NoronhaHostler:2007fg,long}.  This comes about because Hagedorn states allow for very short chemical equilibration times, which were previously not seen using only binary collisions.  Previously, it was assumed that anti-baryons and anti-hyperons were already in chemical equilibrium following the phase transition from QGP to the hadron gas phase because gluon fusion more efficiently produces strange quarks.  However, reactions including Hagedorn states may also reproduce anti-baryons and anti-hyperons within the lifetime of a fireball.

Because of the importance of Hagedorn states close to $T_c$ it is possible that they would have an effect on thermal fits, which are computed within statistical models to reproduce hadron yield ratios in heavy ion collisions \cite{thermalmodels,StatModel,Schenke:2003mj,RHIC,Andronic:2005yp,Manninen:2008mg}.   For Au+Au collisions at RHIC$\;$ at $\sqrt{s_{NN}}=200$ GeV, specifically, estimates for the chemical freeze-out temperature and baryon chemical potential range from $T_{ch}=155-169$ MeV and $\mu_b=20-30$ MeV \cite{RHIC,Andronic:2005yp,Manninen:2008mg}.  Moreover, because Hagedorn states are dependent on the limiting Hagedorn temperature $T_H=T_c$, a relationship between the chemical freeze-out temperature and the critical temperature can be found by including Hagedorn states in thermal fits \cite{NoronhaHostler:2009tz}. This uniquely gives us the ability to distinguish between different critical temperature regions depending on the quality of the fit obtained using the statistical model. Thus, in this paper we explore the possibility of including Hagedorn states into thermal fits.  

\section{Setup}

In this paper we use a grand-canonical model to describe the particle densities from which we can calculate the corresponding ratios as described in detail in \cite{StatModel}.  Hagedorn states are included in our hadron resonance gas model via the exponentially increasing density of states \cite{Hagedorn:1968jf}
\begin{equation}\label{eqn:fitrho}
    \rho(M)=\int_{M_{0}}^{M}\frac{A}{\left[m^2 +m_{r}^2\right]^{\frac{5}{4}}}e^{\frac{m}{T_{H}}}dm,
\end{equation}
where $M_{0}=2$ GeV and $m_{r}^2=0.5$ GeV.  We consider  two different lattice results for $T_c$: $T_c=196$ MeV \cite{Cheng:2007jq,Bazavov:2009zn} (the corresponding fit to the trace anomaly is then $A=0.5 GeV^{3/2}$, $M=12$ GeV, and $B=\left(340 MeV\right)^4$), which uses an almost physical pion mass, and $T_c=176$ MeV \cite{zodor} (the corresponding fit to the energy density leads to $A=0.1 GeV^{3/2}$, $M=12$ GeV, and $B=\left(300 MeV\right)^4$). Both are shown and discussed in \cite{long}. Furthermore, we take into account repulsive interactions using volume corrections  \cite{NoronhaHostler:2008ju,long,Kapusta:1982qd} that are thermodynamically consistent. Note that $B$ is a free parameter  based upon the idea of the MIT bag constant. Also, we include all the known hadrons below with mass below 2 GeV in our analysis.  Note that $B$ is a free parameter  based upon the idea of the MIT bag constant.  Note that due to the volume corrections $T_H>T_c$ for a discussion on this see \cite{long,Meyer:2009tq}.

In our model we do not just consider the direct number of hadrons but also the indirect number that comes from other resonances.  For example, for pions we consider also the contribution from resonances such as $\rho$'s, $\omega$'s, and etc.  The number of indirect hadrons can be calculated from the branching ratios for each individual species in the particle data book \cite{Eidelman:2004wy}. Moreover, there is also a contribution from the Hagedorn states to the total number of pions, kaons, and so on as described in \cite{NoronhaHostler:2007jf,long}.  Thus the total number of ``effective" pions can be described by
\begin{eqnarray}\label{eqn:effpi}
\tilde{N}_{\pi}&=&N_{\pi}+\sum_{i}N_{i}\langle n_{i}\rangle
\end{eqnarray}
whereas the total number of ``effective" protons, kaons, or lambdas (generalized as $X$) can be described by
\begin{eqnarray}\label{eqn:effbbkk}
\tilde{N}_{X}&=&N_{X}+\sum_{i}N_{i}\langle X_{i}\rangle
\end{eqnarray}
where $\langle X\rangle$ is the average number of $X=$ p's, K's, or $\Lambda$'s.
Here $N$ is the total number of each species and $\langle n_{i}\rangle $ is the average number of pions that each Hagedorn state decays into.  To determine $\langle X\rangle$ we use the multiplicities in Fig.\ 2 of Ref.\ \cite{Greiner:2004vm} from the microcanonical model in \cite{Liu} such that
\begin{eqnarray}\label{eqn:gamfit}
  p&=& 0.058\;m_{i}-0.10 \nonumber\\
   K^{+}&=&0.075\;m_{i}+0.047\nonumber\\
  \Lambda&=&0.04\;m_{i}-0.07.
\end{eqnarray}
Clearly, they are all dependent on the mass of the $i^{th}$ Hagedorn state. 

In order to get an idea of the quality of the thermal fits, we define $\chi^2$ as
\begin{equation}
\chi^2=\sum_i \frac{\left(R_i^{exp}-R_i^{therm}\right)^2}{\sigma_i^2}
\end{equation}
where $R_i^{therm}$ is our ratio of hadron yields calculated within our thermal model whereas $R_i^{exp}$ is the experimentally measured value of the hadron yield with its corresponding error $\sigma_i^2$.  Then, $\mu_b$ and $T_{ch}$ are varried until the minimum $\chi^2$ is found. We use the experimental data from both STAR \cite{STAR} and PHENIX \cite{PHENIX}  at mid-rapidity for Au+Au collisions at RHIC at $\sqrt{s_{NN}}$=200 GeV.  Specifically, we observe the ratios: $\pi^{-}/\pi^{+}$, $\bar{p}/p$, $K^-/K^+$, $K^+/\pi^+$, $p/\pi^+$, and $(\Lambda+\bar{\Lambda})/\pi^+$.  All of which are calculated by STAR \cite{STAR}.  However, only $\pi^{-}/\pi^{+}$, $\bar{p}/p$, $K^-/K^+$, $K^+/\pi^+$, $p/\pi^+$ are given by PHENIX.  Because there is such a significant difference between $p/\pi^+$ from PHENIX and STAR we choose only the value from STAR so that we can compare are results to \cite{Andronic:2005yp} where they also exclude $p/\pi^+$ from PHENIX.  It should be noted that Ref.\ \cite{Andronic:2005yp} includes more ratios than we do such as multi-strange particles and resonances, which are not included in this paper.  This is because the purpose of this paper is not to confirm their results, which have already been confirmed in \cite{Manninen:2008mg}, but rather to compare thermal fits that include the contribution of Hagedorn states and those that exclude them.

\section{Results}

\begin{figure}[h]
\centering
\includegraphics[width=7cm]{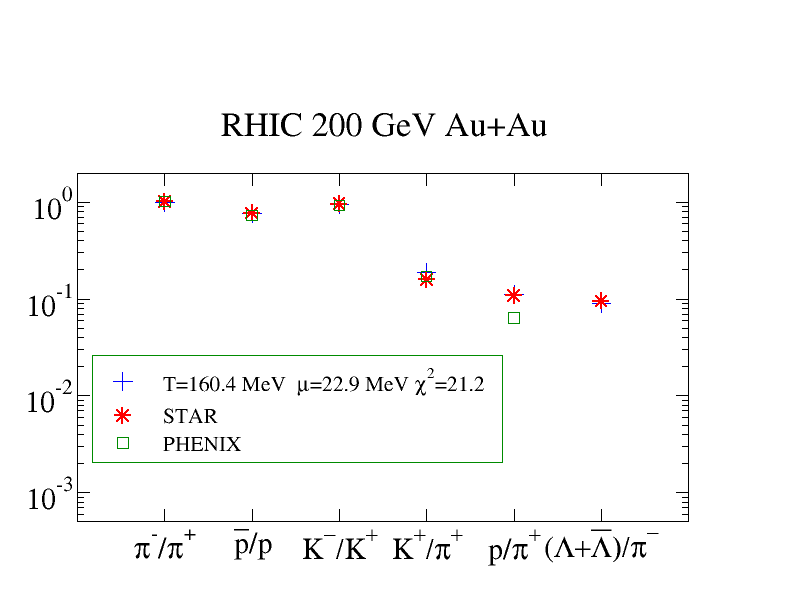} 
\caption{Thermal fits for Au+Au collisions at RHIC at $\sqrt{s_{NN}}=200$ GeV without Hagedorn states with mass above 2 GeV.  }
\label{fig:noHS}
\end{figure}

We show the thermal fits for a hadron gas excluding Hagedorn states in Fig.\ \ref{fig:noHS}, where  $T_{ch}=160.4$ MeV, $\mu_b=22.9$ MeV, and $\chi^2=21.2$.  This is almost identical  to \cite{Andronic:2005yp} where $T_{ch}=160.5$ and $\mu_b=20$ MeV.

\begin{figure}[h]
\centering
\includegraphics[width=7cm]{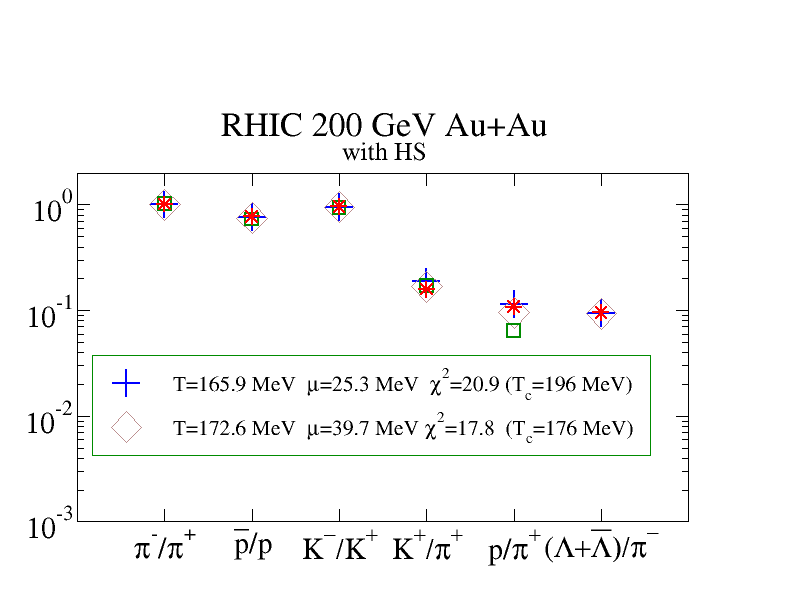} 
\caption{Thermal fits including Hagedorn states with mass above 2 GeV for Au+Au collisions at RHIC at $\sqrt{s_{NN}}=200$ GeV.  }
\label{fig:HS}
\end{figure}

The inclusion of Hagedorn states is our primary interest.  Starting with the fit for the RBC-Bielefeld collaboration, we obtain $T_{ch}=165.9$ MeV, $\mu_b=25.3$ MeV, and $\chi^2=20.9$, which is shown in Fig.\ \ref{fig:HS}.  The $\chi^2$ is actually slightly smaller than in Fig.\ \ref{fig:noHS}. 
When we consider the lattice results from BMW, which are at the lower end of the critical temperature spectrum where $T_c=176$ MeV, we find $T_{ch}=172.6$ MeV, $\mu_b=39.7$ MeV, and $\chi^2=17.8$. The lower critical temperature seems to have a significant impact on the thermal fit. The lower $\chi^2$ is due to the larger contribution of Hagedorn states at at $T_{ch}=172.6$ MeV, which is much closer to $T_c$.  

The difference in the $\chi^2$'s for BMW and RBC-Bielefeld collaboration is directly related to the contribution of Hagedorn states in the model.  Because the RBC-Bielefeld critical temperature region is significantly higher than its corresponding chemical freeze-out temperature the contribution of the Hagedorn states is minimal $\sim$ 4-11$\%$ whereas the contribution from the Hagedorn states is $30-40\%$ for $T_H=176$ MeV (see \cite{NoronhaHostler:2009tz}).  We find that the inclusion of Hagedorn states should not only provide a better fit but they also affect the chemical freeze-out temperature and the baryonic chemical potential. The more mesonic Hagedorn states are present the larger $\mu_b$ becomes.  Furthermore, our fits also have higher $T_{ch}$'s than seen in the fit without the effects of Hagedorn states.

\section{Conclusion}

Our results for thermal fits without Hagedorn states concur well with other thermal fit models \cite{Andronic:2005yp} where the chemical freeze-out temperature ($T_{ch}=160.4$ MeV)  is almost identical and the baryonic chemical potential ($\mu_b=22.9$ MeV)  is only slightly larger. The thermal fit with the known particles in the particle data group provides a decent fit with $\chi^2=21.2$.  However, the inclusion of Hagedorn states provides an even better fit to the experimental data. In fact, we find $\chi^2=17.8$, $T_{ch}=172.6$ MeV, and $\mu_b=39.7$ MeV for the BMW collaboration while for the RBC-Bielefeld collaboration we obtained $\chi^2=20.9$, $T_{ch}=165.9$ MeV, and $\mu_b=20.9$ MeV. This provides further evidence \cite{NoronhaHostler:2008ju,NoronhaHostler:2007jf,NoronhaHostler:2007fg,long} that Hagedorn states should be included in a description of hadronic matter near $T_c$. Since the chemical freeze-out temperature was found to increase from $160$ MeV to roughly $165$ MeV (RBC-Bielefeld) or $172$ MeV (BMW) when including Hagedorn states, this exemplifies the degree of uncertainty in extracting chemical freeze-out thermodynamical parameters by means of such thermal analyzes.

\section{Acknowledgements}

This work was supported by the Helmholtz International
Center for FAIR within the framework of the
LOEWE program (Landes-Offensive zur Entwicklung
Wissenschaftlich-\"okonomischer Exzellenz) launched by
the State of Hesse. J.N. was supported by the US-DOE Nuclear Science Grant No.\ DE-FG02-93ER40764.

\section*{References}
%%%%%%%%%%%%%%%%%%%%%%%%%%%%%%%%%%%%%%%%%%%%%%%%%%%%%%%%%%%%%%%%%%%%%%%

%%%%%%%%%%%%%%%%%%%%%%%%%%%%%%%%%%%%%%%%%%%%%%%%%%%%%%%%%%%%%%%%%%%%%%%

\end{document}